\documentclass[aps,prd,twocolumn,showpacs,superscriptaddress,preprintnumbers,amsmath,amssymb,floatfix]{revtex4}

\usepackage{color}
\usepackage[dvips]{graphicx}

\newcommand{\lsim}{\mathrel{\mathop{\kern 0pt \rlap {\raise.2ex\hbox{$<$}}}\lower.9ex\hbox{\kern-.190em $
\sim$}}}

\newcommand{\gsim}{\mathrel{\mathop{\kern 0pt \rlap{\raise.2ex\hbox{$>$}}}\lower.9ex\hbox{\kern-.190em $\sim
$}}}
\newcommand{\beq}{\begin{equation}}
\newcommand{\eeq}{\end{equation}}

\newcommand{\be}{\begin{equation}}
\newcommand{\ee}{\end{equation}}
\newcommand{\beqarr}{\begin{eqnarray}}
\newcommand{\eeqarr}{\end{eqnarray}}

\begin{document}

\preprint{DFTT 72/2009}

\title{Relic neutralinos and the two dark matter candidate events \\
of the CDMS II experiment}


\author{A. Bottino}
\affiliation{Dipartimento di Fisica Teorica, Universit\`a di Torino, via P. Giuria 1, I--10125 Torino, Italy}
\affiliation{Istituto Nazionale di Fisica Nucleare, via P. Giuria 1, I--10125 Torino, Italy}
\author{F. Donato}
\affiliation{Dipartimento di Fisica Teorica, Universit\`a di Torino, via P. Giuria 1, I--10125 Torino, Italy}
\affiliation{Istituto Nazionale di Fisica Nucleare, via P. Giuria 1, I--10125 Torino, Italy}
\author{N. Fornengo}
\affiliation{Dipartimento di Fisica Teorica, Universit\`a di Torino, via P. Giuria 1, I--10125 Torino, Italy}
\affiliation{Istituto Nazionale di Fisica Nucleare, via P. Giuria 1, I--10125 Torino, Italy}
\author{S. Scopel}
\affiliation{School of Physics and Astronomy, Seoul National University\\
 Gwanak-ro 599, Gwangak-gu, Seoul 151-749, Korea}

\date{December 20, 2009}

\begin{abstract}
The CDMS Collaboration has presented its results for the final exposure of the CDMS II experiment and reports that two candidate events for dark matter would survive after application of the various discrimination and subtraction procedures inherent in their analysis. We show that a population of relic neutralinos, which was already proved to fit the DAMA/LIBRA data on the annual modulation effect, could naturally also explain the two candidate CDMS II events, if these are actually due to a dark matter signal.
\end{abstract}


\maketitle


\begin{figure}[t]
\includegraphics[width=1.1\columnwidth]{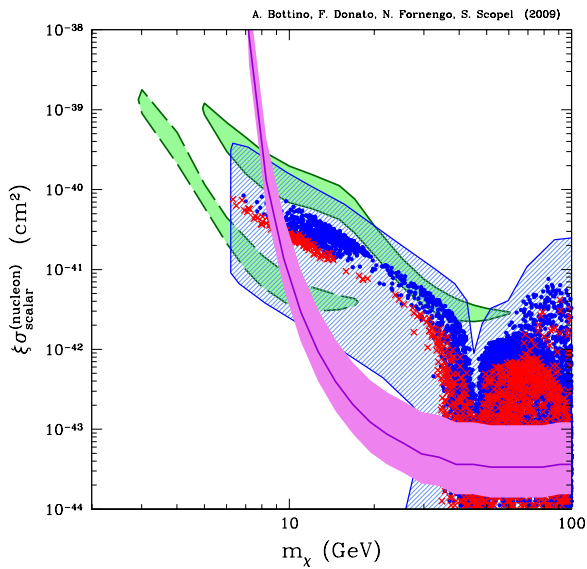}
\vspace{-3em}
\caption{$\xi \sigma_{\rm scalar}^{(\rm nucleon)}$
 as a function of the WIMP mass. The
(green) shaded regions denote the DAMA/LIBRA \cite{dama08} annual modulation
regions, under the hypothesis that the effect is due to a WIMP
with a coherent interaction with nuclei; the region delimitated
by the solid line refers to the case where the channeling effect 
is not included, the one with a dashed contour to the
case where the channeling effect is included \cite{direct}. The (violet)
band displays the region related to the two CDMS candidate
events, obtained from the total rate in the whole energy window. 
The scatter plot represents supersymmetric 
configurations calculated with the model summarized in the Appendix,
at a fixed representative set of values for the hadronic quantities. 
The (red) crosses denote configurations with a neutralino 
relic abundance which matches the WMAP cold dark
matter amount (0.098 $\leq \Omega_{\chi} h^2 \leq$ 0.122), 
while the (blue) dots
refer to configurations where the neutralino is subdominant
($ \Omega_{\chi} h^2 <$ 0.098). The region covered by 
a (blue) slant hatching denotes the extension of the scatter plot 
upwards and downwards, when the hadronic uncertainties in the
scattering coherent cross--section are included.}
\label{fig:CDMS2.ps}
\end{figure}

\begin{figure}[t]
\includegraphics[width=1.1\columnwidth]{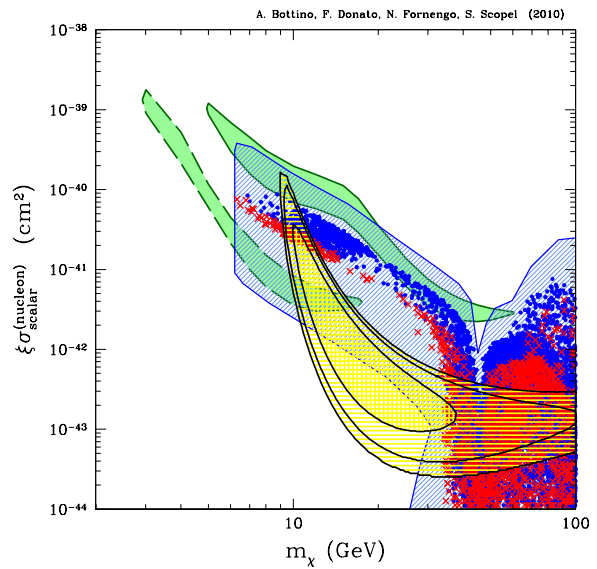}
\vspace{-3em}
\caption{The same as in Fig. \ref{fig:CDMS2.ps}, except that the
(yellow) shaded regions compatible with the CDMS II candidate events are obtained by
a maximal likelihood method applied to the differential energy recoil rate, under the
hypothesis of negligible background.
The contours refer to (from the internal to the external one) 68\%, 90\%
and 95\% C.L.
}
\label{fig:CDMS2_like_noback.ps}
\end{figure}

\begin{figure}[t]
\includegraphics[width=1.1\columnwidth]{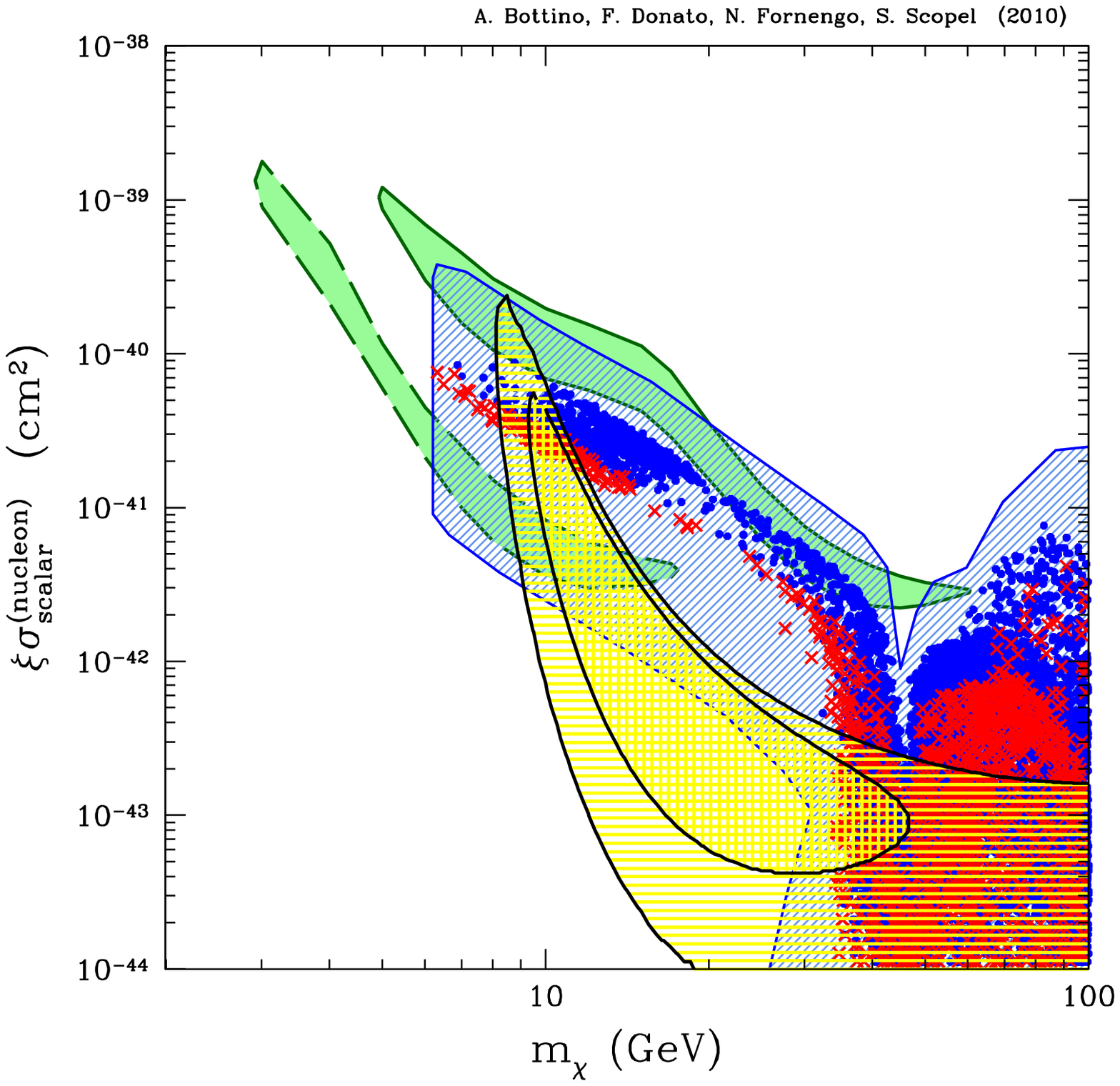}
\vspace{-3em}
\caption{The same as in Fig. \ref{fig:CDMS2_like_noback.ps}, but under the hypothesis
of a backgound contribution as in Ref. \cite{Kopp:2009qt}, normalized to 0.8 events in
the whole energy window of CDMS II. The contours refer to (from the internal to the external one) 68\% and 85\% C.L.}
\label{fig:CDMS2_like_back.ps}
\end{figure}

The search for a sign from Dark Matter (DM) involves
direct detection, consisting in the measurement of the effects induced by
the feeble interaction of the DM particles with the material
of a low--background  set--up, and indirect measurements. These concern
many possible signals, ranging from neutrinos to
charged cosmic rays (positrons, antiprotons, antideuterons),
to gamma rays, to a radio signal and even to effects induced on the cosmic
microwave background.

In Ref. \cite{direct} we have shown that  the annual modulation effect  at a
8.2 $\sigma$ C.L.,
 obtained by the DAMA/NaI and DAMA/LIBRA experiments (with a total exposure of 0.82 ton yr) \cite{dama08}
 is very well fitted by relic neutralinos
in  an effective Minimal Supersymmetric Standard Model (effMSSM) at the electroweak scale defined in terms
of a limited number of parameters. We recall that
the effect measured by the DAMA Collaboration is  the first and up--to--now unique evidence for a signal compatible with a
typical signature (annual modulation) expected for dark matter particles \cite{freese}.

Other experiments of WIMP direct detection do not currently have the capability of measuring
 the annual modulation effect and usually provide upper bounds for the expected
 signals \cite{others}. These limits are obtained through complex
 procedures for discriminating electromagnetic signals from recoil events and through delicate
 subtractions meant to separate putative WIMP signals from neutron--induced events. A major
 critical point in these experiments and related analyses is that the very signature
 (the annual modulation) of the searched signal cannot be employed in extracting the authentic events.
Other potential difficulties are related to stability features and determination of the threshold and of the energy scale.

In Ref. \cite{direct} it was also pointed out that
 the inclusion of these upper bounds,  taken at their face value,
would  anyway allow a compatibility with the DAMA data for  a  range in the
WIMP (neutralino) mass around 7--10 GeV. A similar result has been obtained also in Ref. \cite{Savage:2008er}.

The CDMS Collaboration has now presented its results for the final exposure of the CDMS II experiment \cite{cdms}. In that paper it is reported that two candidate events for DM would survive after application of various discrimination and subtraction procedures, though a probability of 23\% exists  that they are of a more prosaic origin. These 2 events have recoil energies of 12.3 keV and 15.5 keV. 

If one assumes that the two candidate events are due to a WIMP particle
with a coherent interaction with nuclei, taking into account the CDMS
total exposure, one can derive that the relevant 90\% C.L. region in the
plane $m_{\chi}$--$\xi \sigma_{\rm scalar}^{(\rm nucleon)}$ (up to a
WIMP mass of 100 GeV) is the one displayed in Fig. 1 ($m_{\chi}$ denotes
the WIMP mass, $\sigma_{\rm scalar}^{(\rm nucleon)}$ is the
WIMP--nucleon coherent cross section and $\xi$ is the WIMP local
fractional density). In this Figure, due to the low statistics,
we have adopted the simple criterion
to require $n=2$ WIMP events ($0.6<n<4.7$ at 90\% C.L. 
for a Poissonian distribution) in the total
range of the recoil energy $E_{R}$ observed by CDMS, $10~\mbox{keV} <
E_{R}< 100~\mbox{keV}$. This is sufficient to capture the main features of the allowed
region. This is also true for $m_{\chi}\lsim 8.5-9$ GeV where, depending
on the values of the escape velocity in the Galaxy and on the rotational
velocity of the Solar System, the event with $E_{R}=15.5$ keV could
in principle not be ascribed to a WIMP. In fact in this case the region
shown in Fig. 1 overlaps with the one (not shown in Figure \ref{fig:CDMS2.ps}) obtained by requiring only one WIMP
event ($0.11 < n < 3.44$ at 90\% C.L.) for $10~\mbox{keV}<E_{R}<12.3~\mbox{keV}$:
the region obtained with this criterion has the upper boundary
about 25\% smaller than the one shown in Fig. \ref{fig:CDMS2.ps} and the lower boundary
reduced by about a factor of 5.

In Figure \ref{fig:CDMS2.ps} also the annual modulation regions  of the DAMA Collaboration are shown, with and without inclusion of the channeling effect \cite{channeling}. The exact modeling of channeling is still under study, then one expects that the actual physical situation is comprised within the two regions represented in the figure. As a reference model for the WIMP halo distribution, a cored--isothermal sphere is employed with the following parameters:
local value of the rotational velocity $v_0 = 220$ km s$^{-1}$, escape velocity $v_{\rm esc}=650$ km s$^{-1}$
and total non-baryonic dark--matter density $\rho_0 = 0.34$ GeV cm$^{-1}$. Obviously, the DM halo 
distribution could be quite different \cite{bcfs}: this would induce a shift of the actual position of the regions and bounds, as discussed and shown e.g. in Ref. \cite{direct}.

In Fig. 1 we also display the scatter plot representing the supersymmetric configurations
calculated within a realization of the Minimal Supersymmetric Standard Model where 
gaugino unification is relaxed \cite{light,direct}. For convenience, the model is summarized in the Appendix. The scatter plot refers to
a fixed representative set of values for the hadronic
quantities involved in the neutralino--nucleon cross sections
\cite{direct}.
The region covered by a (blue) slant hatching denotes the extension of the scatter plot upwards and
downwards, when the hadronic uncertainties extensively
discussed in Ref. \cite{direct} are included.

From Fig. \ref{fig:CDMS2.ps} and the previous discussion about the CDMS region one finds that the putative
CDMS events are compatible simultaneously with the
DAMA/LIBRA data and the theoretical evaluations in
the mass range 8--12 GeV. 
It is worthwile to point out that at such low masses the expected recoil spectrum depends on the high velocity tail of the velocity distribution, which is sensitive to the details of the astrophysical
model and in particular to the escape velocity. Other
possibilities for the modeling of the velocity distribution have been
discussed in \cite{bcfs}.
We also stress that
the explanation -- in terms of the relic neutralinos in the supersymmetric model discussed here -- 
of the annual--modulation data alone extends over a 
much wider range; for instance, for the case
of a WIMP halo distribution given by a cored--isothermal
sphere with the parameters mentioned above, this extended 
range can be simply read from Fig. 1 to be 6
GeV $\lsim m_{\chi} \lsim$ 60 GeV. These light neutralinos can also
be complementarily investigated by indirect means, such
as cosmic antiprotons \cite{direct,antip} and antideuterons \cite{antid,direct}, signals at neutrino telescopes
\cite{neutr}, and, most notably, can be searched for at the Large
Hadron Collider \cite{lhc}. 
Astrophysical bounds arising from multi--wavelength analyses  \cite{cosmo}, which may
be strong depending on assumptions on the DM distribution and on astrophysical properties, 
like
those related to cosmic--rays propagation and energy losses, do not markedly constrain 
the supersymmetric configurations of Fig. \ref{fig:CDMS2.ps},
especially when astrophysical uncertainties are properly taken into account.

Our previous analysis was based only on the
total rate taken over the whole recoil energy range observed by CDMS II,
without using any spectral information. This is motivated by the very low statistics (2 events), which makes very critical (and to some extent not fully justified) a statistical analysis of the energy spectrum. However, forcing somewhat the situation, one can wonder what would produce an analysis in terms of the energy spectrum. In Fig. \ref{fig:CDMS2_like_noback.ps} and \ref{fig:CDMS2_like_back.ps} we therefore
show the regions compatible with the 2 CDMS II events at 12.3 keV and 15.5 keV, taking into account the spectral behaviour of the theoretical recoil rate. In the determination of the allowed regions we have adopted a maximal likelhood analysis \cite{PDG}. Fig. \ref{fig:CDMS2_like_noback.ps} shows the case of a negligible background contribution, and the
contours refer to a confidence level of 68\%, 90\% and 95\%, from the innermost to
the outermost. Fig. \ref{fig:CDMS2_like_back.ps} instead refers to the presence of a 
background contribution, which we have modeled as in Ref. \cite{Kopp:2009qt}, {\em i.e.}
with an energy dependence $dN/dE = -0.00295 + 0.463/E$ normalized to the total number of event of 0.8, to conform to an estimate of the background contribution of
$0.8 \pm 0.1 \mbox{(stat)} \pm 0.2 \mbox{(syst)}$\cite{cdms}. In the case
of Fig. \ref{fig:CDMS2_like_back.ps}, the contours refer to 68\% and 85\% C.L.,
and they evolve into an open region  ({\em i.e.} into an upper bound) at the 90\% C.L., a result compatible to the one
obtained in Ref. \cite{Kopp:2009qt}, where a slightly different statistical analysis
is adopted.
Fig. \ref{fig:CDMS2_like_noback.ps} and \ref{fig:CDMS2_like_back.ps} are quite
compatible with the results of Fig. \ref{fig:CDMS2.ps} obtained by using the total
counting number, and reinforce our conclusions of compatibility between DAMA and CDMS II
for light WIMPS, and between these experimental results and our SUSY models
with light neutralino dark matter.

In conclusion, in this note we have considered the two
events which, in the analysis of CDMS II, seem to survive 
after the various discrimination and subtraction proce-
dures. We have shown that, should these events be due
to WIMP--nucleus coherent interactions, this result would
be compatible both with the annual--modulation signal previously reported  by the DAMA Collaboration and
with an interpretation in terms of relic light neutralinos.
This conclusion is not affected by other upper bounds of direct dark
matter detection. In particular, the XENON upper bound \cite{aprile} suffers from
large uncertaities due to conflicting determinations of the
scintillation efficiency at low nuclear recoils (as shown in Fig. 12 of
\cite{mazur}). ÊCalculations performed in Ref. \cite{mazur} (though with a threshold
energy somewhat smaller than the one of XENON10) indicate that at a WIMP
mass of order 10 GeV the bound of Ref. \cite{aprile} should be relaxed by more
than an order of magnitude. One should furthermore note that in XENON10
the energy scale is particularly uncertain due to a calibration at an
energy much higher than the declared threshold energy.

\section{Appendix: The supersymmetric model}
\label{sec:susy}

The supersymmetric scheme we employ in the present paper is the
one described in Ref. \cite{direct} as an effective
 Minimal Supersymmetric Standard Model (effMSSM) at the electroweak scale, with the following independent
parameters: $M_1, M_2, M_3, \mu, \tan\beta, m_A, m_{\tilde q}, m_{\tilde l}$
and $A$. Notations are as
follows: $M_1$,  $M_2$  and $M_3$ are the U(1), SU(2) and SU(3)  gaugino masses
(these parameters are taken here to be positive),
$\mu$ is the Higgs mixing mass parameter, $\tan\beta$ the
ratio of the two Higgs v.e.v.'s, $m_A$ the mass of the CP-odd
neutral Higgs boson, $m_{\tilde q}$ is a squark soft--mass common
to all squarks, $m_{\tilde l}$ is a slepton soft--mass common to
all sleptons, and $A$ is a common dimensionless trilinear parameter
for the third family, $A_{\tilde b} = A_{\tilde t} \equiv A
m_{\tilde q}$ and $A_{\tilde \tau} \equiv A m_{\tilde l}$ (the
trilinear parameters for the other families being set equal to
zero). In this model no gaugino mass unification at a Grand Unified
scale is assumed, whence light neutralinos arise.

The parameter space of this model is bounded by a large host
of experimental data: invisible Z decay (for decay into light neutralinos),
direct searches of supersymmetric particles and higgs bosons at LEP and Tevatron,
supersymmetric constributions to rare processes: 
$BR(b \rightarrow s + \gamma)$, $BR(B_s^{0} \rightarrow \mu^{-} + \mu^{+})$,
measurements of the muon anomalous magnetic moment $a_\mu \equiv (g_{\mu} -
2)/2$. Other details about the model and the relevant constraints can be found in
Ref. \cite{direct}. 

\section{Note added}
After submission of the present paper, the CoGeNT Collaboration
\cite{cogent} has presented the results of a search for light--mass DM
particles, where an irreducible excess of bulk--like events below an
energy of 3 keV is observed. As discussed in Ref. \cite{cogent}, should this
population of events be due to WIMP interactions with the detector,
these would entail a WIMP mass of 7--12 GeV with a cross section
$m_{\chi}$--$\xi \sigma_{\rm scalar}^{(\rm nucleon)} \simeq (3-10) \times
10^{-41}$ cm$^2$, thus in a region in Êagreement Êwith predictions of
our model, with the DAMA/LIBRA and CDMS II results.

\acknowledgments

Work supported by research grants funded jointly by Ministero dell'Istruzione,
dell'Universit\`a e della Ricerca (MIUR, under constract number 2008NR3EBK), by Universit\`a di Torino, by
Universit\`a di Padova, by the Istituto Nazionale di Fisica Nucleare (INFN) 
within the {\sl Astroparticle Physics Project}. SS is supported by the WCU 
program (R32-2008-000-10155-0) of National Research Foundation of Korea.
NF acknowledges support of the spanish MICINN ConsoliderÐIngenio 2010 Programme under
grant MULTIDARK CSD2009-00064.


\end{document}